\documentclass[11pt]{article}
\usepackage{moriond,epsfig}

\def\Journal#1#2#3#4{{#1} {\bf #2}, #3 (#4)}

\def\PLB{{\em Phys. Lett.}  B}

\def\PRD{{\em Phys. Rev.} D}

\def\EPJ{{\em Eur. Phys. J.} C}

\def\be{\begin{equation}}
\def\ee{\end{equation}}
\def\bea{\begin{eqnarray}}
\def\eea{\end{eqnarray}}

\begin{document}
\vspace*{4cm}
\title{Recent NA48 results on rare K decays}

\author{ C. Cheshkov }

\address{DSM/DAPNIA - CEA Saclay, F-91191 Gif-sur-Yvette, France\\
FOR THE NA48 COLLABORATION}

\maketitle\abstracts{
Recent NA48 results from detailed studies of $K_{L,S}\to \pi^+ \pi^- e^+ e^-$
and $K_S\to\pi^0 \gamma\gamma$ decay modes are presented. The results are based on the data collected with the NA48 detector at the CERN SPS during the 1998-1999 and 2000 data taking periods, respectively.
Prospects for future results on charged kaon decays are briefly described.}

\section{NA48 experiment}

The NA48 experiment has been designed for a precise measurement of the
CP-violation parameter $Re(\epsilon'/\epsilon)$. It makes use of simultaneous
$K_L$ and $K_S$ beams produced at two different targets, situated 120m from
each other. The main two elements of the setup are charged particle magnetic
spectrometer and liquid krypton electromagnetic calorimeter (LKr).
The spectrometer consists of a dipole magnet with a horizontal
transverse momentum kick of 256 MeV/c and a set of four drift chambers
(two upstream of the magnet and two downstream of it). The momentum
resolution of the spectrometer is given by
$\sigma_P/P(\%)=0.48\oplus 0.009\times P$, where P is in GeV/c.
LKr is a quasi-homogeneous detector, having projective tower structure 
which is formed by copper-beryllium ribbons extending between the front
and the back of the calorimeter with a accordion geometry.
The energy resolution of the calorimeter is
$\sigma_E/E(\%)=3.2\%/\sqrt(E)\oplus10.0/E\oplus0.5$,
where E is given in units of GeV. A detailed description of the whole
NA48 setup can be found elsewhere \cite{eps}.

\section{Data samples}
The $K_L \to \pi^+ \pi^- e^+ e^-$ and a fraction of the
$K_S \to \pi^+ \pi^- e^+ e^-$ data were taken using simultaneous $K_L$
and $K_S$ beams during the 1998 and 1999 SPS
running periods dedicated to the measurement of 
Re($\epsilon'/\epsilon$). The main part of the $K_S$ data was collected
from a short high intensity $K_S$ run in 1999. In both cases
$K_L\to\pi^+\pi^-\pi^0$ mode, followed by the Dalitz decay of $\pi^0$,
was used as a normalization channel.

The data needed to observe and analyze the $K_S\to\pi^0\gamma\gamma$
decay mode was recorded during a part of the 2000 run with high
intensity $K_S$ beam and without the charged particle magnetic
spectrometer. The corresponding branching ratio was determined relative
to the $K_S\to\pi^0\pi^0$ channel.

In order to take into account the detector acceptance and the reconstruction
efficiency, a detailed Monte Carlo program based on GEANT has been employed.

\section{First observation of $K_S\to\pi^0\gamma\gamma$}

As it was recognized in \cite{ecker}, the decay $K_S\to\pi^0\gamma\gamma$ can
provide valuable test of the chiral structure of the weak vertex.
The interest in this particular decay mode is also enhanced by the fact
that it has not been observed so far (recently the NA48 experiment put
the best upper limit for the decay branching ratio to $3.3\times 10^{-7}$
using the 1999 high intensity $K_S$ data).

\begin{figure}
\begin{center}
\epsfig{figure=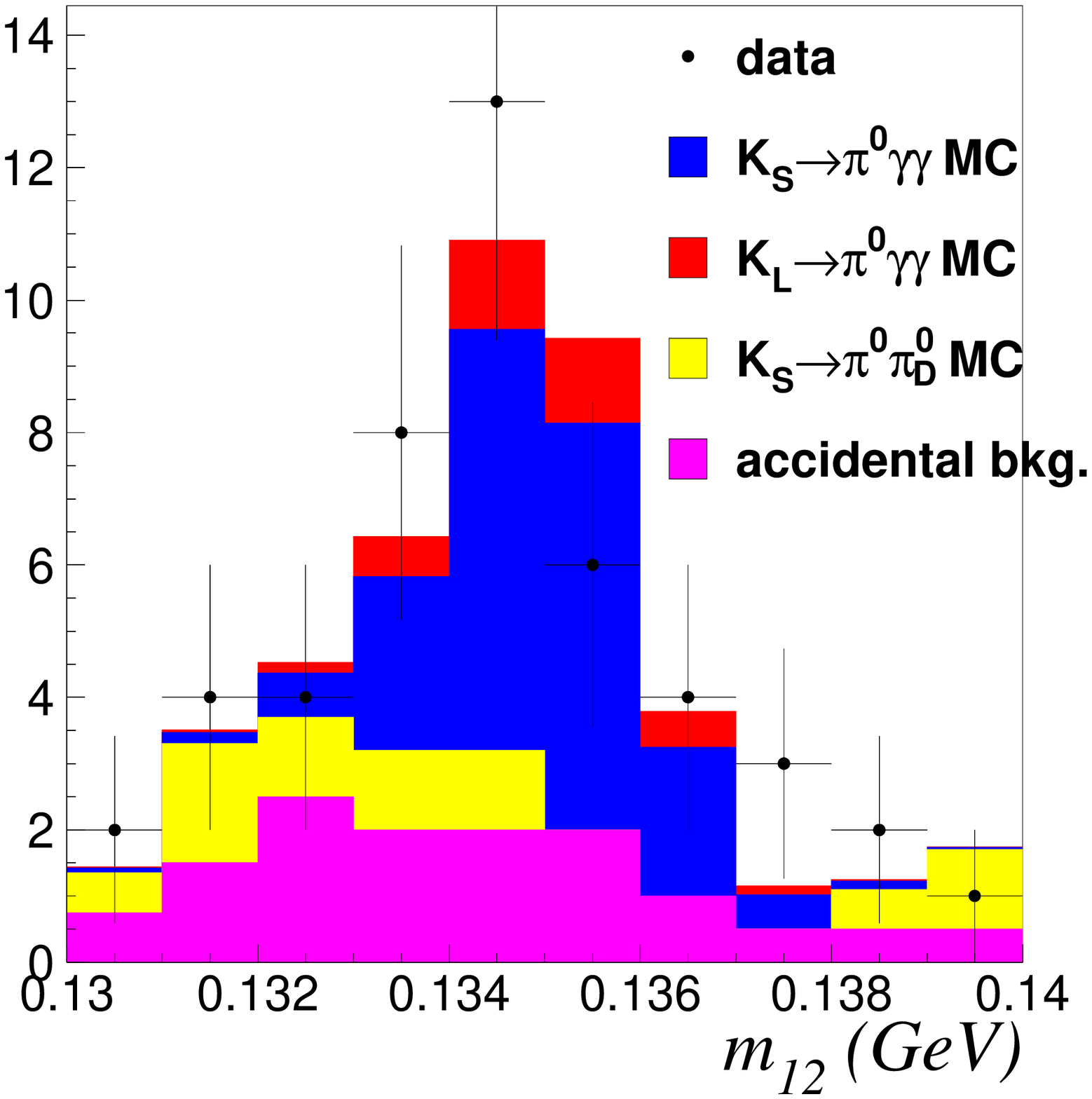,height=6cm}\hspace{3cm}\epsfig{figure=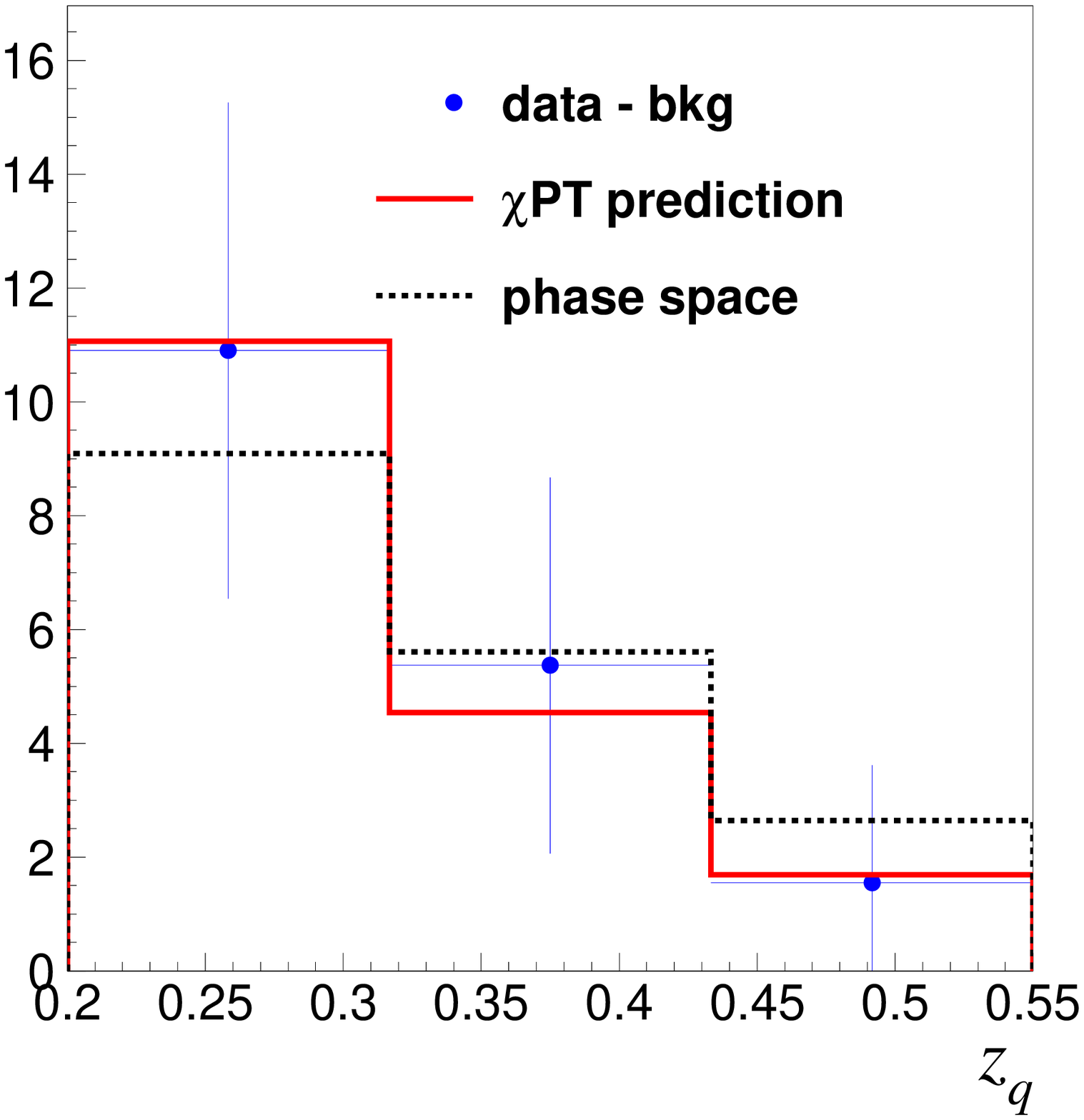,height=6cm}
\caption{$K_S\to\pi^0\gamma\gamma$ candidates: the invariant mass of two photons which form $\pi^0$ (left); the $z=q^2/M_K^2$ distribution (right). 
\label{fig:kspi0gg1}}
\end{center}
\end{figure}

The most important issues in the analysis of $K_S\to\pi^0\gamma\gamma$ are the
efficient rejection and the accurate estimation of the backgrounds.
The following background sources have been identified and carefully studied:
\begin{itemize}
	\item{Accidental beam activity. This background has been rejected by
series of timing cuts and using veto signals from anti-counters which surround
the decay volume. The remaining contribution has been evaluated from the
sidebands of the time distributions.}
	\item{$K_S\to\pi^0\pi^0$ and $K_S\to\pi^0\pi^0$ followed by Dalitz
decay of one of the $\pi^0$s. The suppression of these contributions has been
obtained by applying a set of kinematic cuts against Dalitz decays and 
$K_S\to\pi^0\pi^0$ decays with significant $\gamma$ energy losses.}
	\item{$K_L\to\pi^0\gamma\gamma$. Due to the fact that this background
source is irreducible, its contribution to the signal has been subtracted
using Monte-Carlo generated decays and a value for the flux obtained from
a study of $K_S\to\pi^0\pi^0$ data.}
	\item{$\Xi^0\to\Lambda\pi^0$ followed by $\Lambda\to n\pi^0$. The 
background has been rejected by cuts on the asymmetries between the $\gamma$
 energies. The contribution which survive the cuts has been estimated by 
comparing the shower profiles in the LKr calorimeter for the 
$K\to\pi^0\gamma\gamma$ candidates and reconstructed $\Xi^0$ decays.}
\end{itemize}
The $\pi^0$ invariant mass distribution for $K\to\pi^0\gamma\gamma$
candidates which passed the whole event selection is shown in Figure \ref{fig:kspi0gg1}.
Table \ref{tab:kspi0gg} summarizes the number of events seen inside the
signal region and the estimated contributions from background sources.
The probability that the observed 17.4 signal events are consistent with
background fluctuations is less that $9\times 10^{-4}$ and therefore one
can claim for first observation of this decay mode. The corresponding
branching ratio was found to be
$BR(K_S\to\pi^0\gamma\gamma)|_{z>0.2}=(4.9\pm 1.6_{stat}\pm 0.8_{syst})\times 10^{-8}$, where $z=q^2/M_K^2$, fully consistent with the prediction in \cite{ecker}.
On the other hand,
the available statistics does not allow to test the chiral structure of
the weak vertex (Figure \ref{fig:kspi0gg1}).

\begin{table}[t]
\caption{$K_S\to\pi^0\gamma\gamma$. Summary of the number of events in the signal region and the expected contributions from background sources.\label{tab:kspi0gg}}
\vspace{0.3cm}
\begin{center}
\begin{tabular}{|l|c|c|c|}
\hline
Number of events in signal region & 31.0 & $\pm$ & 5.6\\
\hline
Beam activity                     & -7.4 & $\pm$ & 2.4\\
$K_S\to\pi^0\pi^0_D$              & -2.4 & $\pm$ & 1.2\\
$K_L\to\pi^0\gamma\gamma$         & -3.8 & $\pm$ & 0.0\\
Acceptance                        &      & $\pm$ & 0.7\\
\hline
Number of events after background subtraction & 17.4 & $\pm$ & 6.2\\
\hline
\end{tabular}
\end{center}
\end{table}

\section{Detailed study of $K_{L,S}\to \pi^+ \pi^- e^+ e^-$}

The matrix element of $K_L\to \pi^+ \pi^- e^+ e^-$ decay receives
contributions from CP-violating inner bremsstrahlung, CP-conserving
emission of M1 photon, CP-violating emission of E1 photon and CP-conserving
$K^0$ charge radius processes \cite{sehgal1,sehgal2}. The interference
between two first terms
leads to an observable CP-violating polarization of the virtual $\gamma$.
This observable can be analyzed in terms of asymmetry
$A_{\phi}=(N_{\sin\phi\cos\phi >0}-N_{\sin\phi\cos\phi <0})/(N_{\sin\phi\cos\phi >0}+N_{\sin\phi\cos\phi <0})$, where N represents the number of
observed events and $\phi$ is the angle between the planes formed by
$\pi^+\pi^-$ and $e^+e^-$. Contrary to the $K_L$ case, in the
$K_S\to \pi^+ \pi^- e^+ e^-$ decay the only contribution is the
inner bremsstrahlung.

\begin{figure}
\begin{center}
\epsfig{figure=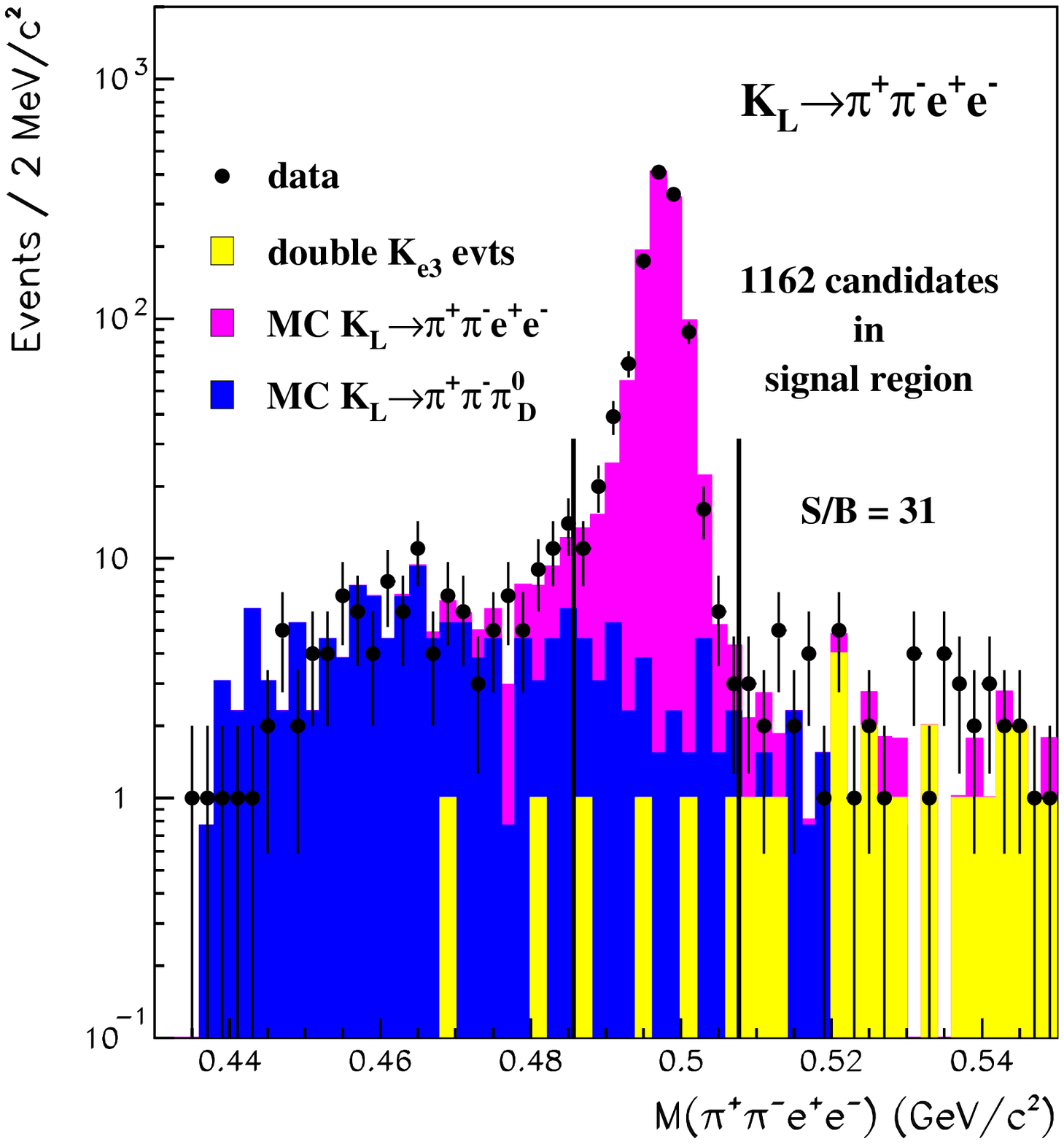,height=6cm}\hspace{1cm}\epsfig{figure=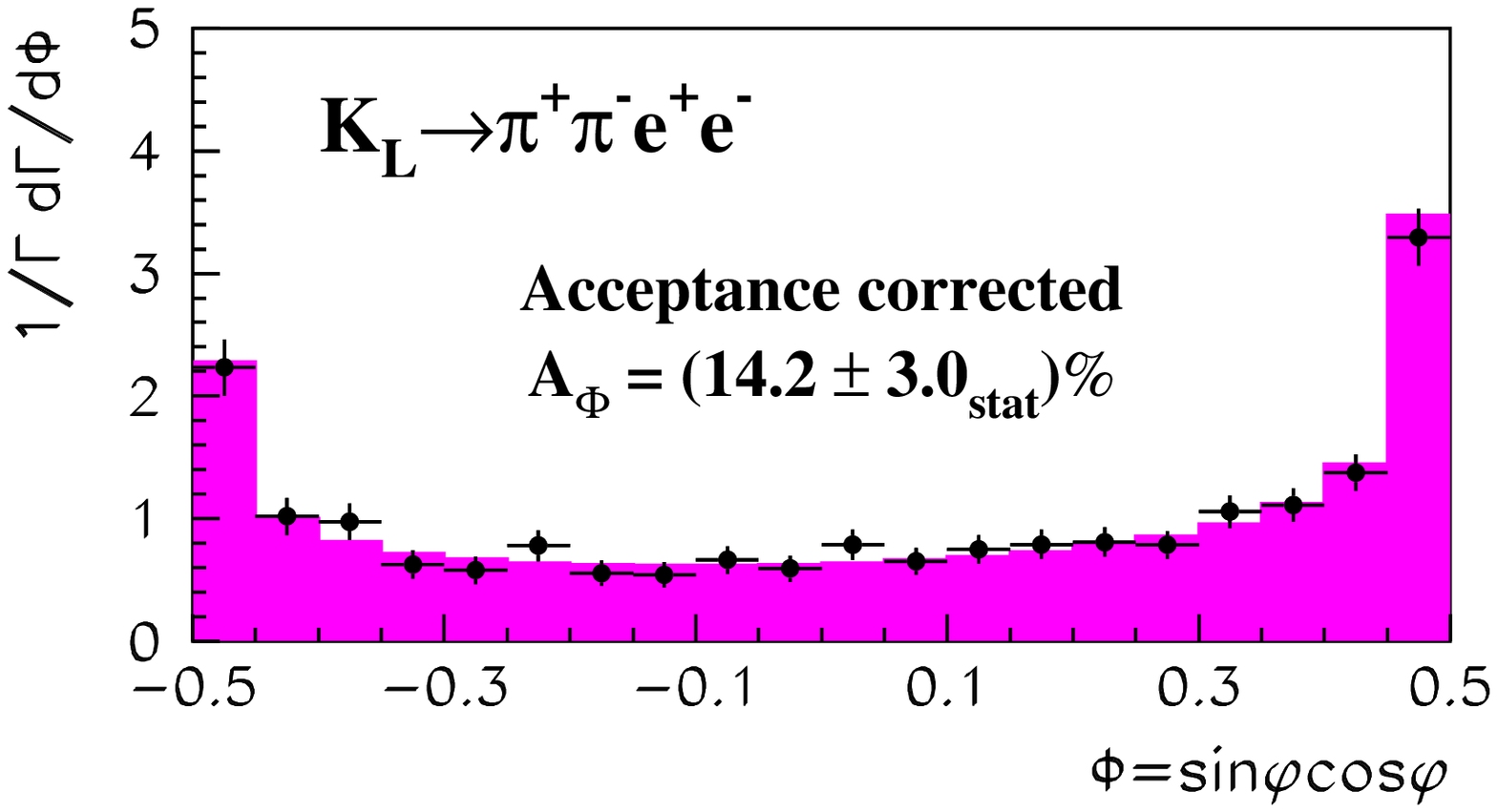,height=6cm,width=8cm}
\caption{$K_L\to\pi^+\pi^- e^+ e^-$: the invariant mass distribution (left); the angular distribution after acceptance correction (right). 
\label{fig:klpipiee1}}
\end{center}
\end{figure}

During the analysis of $K_L\to \pi^+ \pi^- e^+ e^-$ two main background sources
have been identified:
\begin{itemize}
	\item{$K_L\to\pi^+\pi^-\pi^0$ followed by Dalitz decay of $\pi^0$.
The background has been suppressed by strong kinematical cuts against events
with missing particle.}
	\item{Two overlapping in time $K^0e3$ decays. Both the vertex quality
cut and the cut on time difference between $\pi^+ e^-$ and $\pi^- e^+$ pairs
have been applied in order to reduce this kind of background. The remaining
contribution in the signal region has been estimated using accidental events
in which the two pions and two leptons have the same charge.}
\end{itemize}
Figure \ref{fig:klpipiee1} shows the distribution of the invariant mass for
the selected $K_L\to \pi^+ \pi^- e^+ e^-$ candidates. As one can see the
background contamination is small enough allowing an unbiased analysis of
the decay parameters.
In the $K_S$ case the background from $K_L\to\pi^+\pi^-\pi^0_D$ is
significantly lower, but at the same time two new contributions arise - the
irreducible contamination from $K_L\to\pi^+\pi^- e^+ e^-$ and the
$\Xi^0\to\lambda\pi^0_D$ decay.

In order to extract the parameters of the M1 emission and $K^0$
charged radius processes, the selected $K_L\to \pi^+ \pi^- e^+ e^-$
candidates have been fitted to Monte Carlo generated events by a maximum
likelihood method. Similar analysis of the $K_S$ data has confirmed that
it is consistent with pure inner bremsstrahlung contribution.

Based on the observed number of $K_{L,S}\to\pi^+\pi^- e^+ e^-$ events, the corresponding branching ratios has been determined to be
$BR(K_L\to \pi^+ \pi^- e^+ e^-)=(3.08\pm 0.09_{stat}\pm 0.15_{syst}\pm 0.10_{norm})\times 10^{-7}$ and
$BR(K_S\to \pi^+ \pi^- e^+ e^-)=(4.71\pm 0.23_{stat}\pm 0.16_{syst}\pm 0.15_{norm})\times 10^{-5}$, respectively.

Figure \ref{fig:klpipiee1} represents the angular distribution of
$K_L\to \pi^+ \pi^- e^+ e^-$ after a correction for the detector acceptance.
A numerical analysis of this distribution has shown that the asymmetry
$A_\phi=(14.2\pm 3.0_{stat}\pm 1.9_{syst})\%$. The observed $4\sigma$ effect
is a clear indication of CP violation and is in agreement with the theoretical
predictions \cite{sehgal1,sehgal2}. As expected the corresponding $K_S$ asymmetry is compatible with 0.

A detailed description of the $K_{L,S}\to \pi^+ \pi^- e^+ e^-$ analysis and results can be found in \cite{eddy}.

\section{Run 2003}
In 2003 the NA48 collaboration continues its experimental program by moving
to charged kaon sector \cite{na482}. The main goal of the run will be a precise
measurement of the CP-violation in the asymmetry of the Dalitz plots for
$K^+\to (3\pi)^+$ and $K^-\to (3\pi)^-$ decays. Another important aim of
the experiment is to measure the scattering length $a^0_0$ by collecting
about $10^6$ $K^\pm e4$ events. As a by-product it is planned to study also
some rare decay channels like $K^\pm\to\pi^\pm\pi^0\gamma$,
$K^\pm\to\pi^\pm\gamma\gamma$, $K^\pm\to\pi^\pm e^+e^-$ and $K^\pm\to\pi^\pm\pi^0 e^+e^-$. Another interesting possibility will be to extract the CKM matrix
element $V_{US}$ from an analysis of high statistics samples of
$K^\pm e3$ and $K^\pm\mu 3$.

The experimental setup is basically the same with
two major upgrades. The first one is a new beamline configuration which will
provide simultaneous and focused 60 GeV/c $K^\pm$ beams. The second upgrade is
a new beam spectrometer based on MICROMEGAS TPCs. The spectrometer will be able
to work in the extremely high intensity beam environment and will measure the
momentum of the decaying kaons with a precision of 1\%.

\end{document}